\documentclass[aps,prd,a4paper,preprintnumbers,twocolumn,floatfix,showkeys,showpacs,nofootinbib]{revtex4}

\usepackage{psfrag}
\usepackage{amssymb}
\usepackage{subfigure}
\usepackage{color}
\usepackage{mathrsfs}
\usepackage{graphicx}

\def\be{\begin{equation}}
\def\ee{\end{equation}}
\def\bea{\begin{eqnarray}}
\def\eea{\end{eqnarray}}

\begin{document}


\title{Interpreting the conformal cousin of the Husain-Martinez-Nu\~nez spacetime}



\author{Valerio Faraoni}
\email[]{vfaraoni@ubishops.ca}
\affiliation{Physics Department and {\em STAR} Research Cluster, 
Bishop's University, 2600 College Street, 
Sherbrooke, Qu\'ebec, Canada J1M~1Z7}

\author{Andres F. Zambrano Moreno}
\email[]{azambrano07@UBishops.ca}
\affiliation{Physics Department,  
Bishop's University, 2600 College Street, 
Sherbrooke, Qu\'ebec, Canada J1M~1Z7}



\begin{abstract} 
A 2-parameter inhomogeneous cosmology in 
Brans-Dicke theory, obtained by conformally transforming 
the Husain-Martinez-Nu\~nez scalar field solution of 
the Einstein equations is studied and interpreted 
physically. According to the values of the parameters it 
describes a wormhole or a naked singularity. The reasons why 
there isn't a one-to-one correspondence between conformal 
copies of this metric are discussed. 
\end{abstract}


\maketitle




\section{\label{section1}Introduction}

In their low-energy limit, most theories attempting to quantize gravity produce
modifications of general relativity in the form of non-minimally coupled dilaton fields 
and/or higher derivative terms in the gravitational sector (this is the case, for example, 
 of the bosonic string theory which reduces to  an $ \omega= -1 $ Brans-Dicke theory 
\cite{bosonic}).
Attempts to explain the present-day cosmological acceleration discovered using the 
luminosity distance-redshift relation of type Ia 
supernovae \cite{SN} without introducing the {\em ad hoc} dark energy has led, among other 
scenarios, to infrared modifications of gravity \cite{CCT}. 
This ``$f(R)$'' gravity is nothing but a Brans-Dicke theory with a special 
scalar field potential  (see \cite{f(R)reviews} for reviews). 

Several alternative theories of gravity have been proposed and studied recently, as 
low-energy effective actions or as toy models
for quantum or emergent gravity, or in the context of early or late universe cosmology 
(see \cite{CliftonPadillaetc} for a recent review). In addition, varying 
``constants'' of nature hypothesized by Dirac \cite{Dirac}  
can be implemented 
naturally in scalar-tensor gravity, in which the gravitational coupling 
depends on the  spacetime point \cite{BD, ST}. 

When approaching a theory of gravity, it is important to understand its spherically 
symmetric solutions and, in particular, its black holes. Solutions of the field equations 
describing inhomogeneities in cosmological spaces have been studied with the specific 
purpose of modelling spatial variations of the gravitational coupling \cite{varyingG, CMB}.  
Spherically symmetric inhomogeneous solutions of Brans-Dicke gravity which describe a 
central condensation embedded in a Friedmann-Lema\^itre-Robertson-Walker (FLRW) background 
have been found, but not studied or interpreted, in  
Ref.~\cite{CMB}. Extra value is added to 
the study of spacetimes describing central objects in cosmological backgrounds by the fact 
that  such metrics are not well understood even in the context of Einstein theory 
\cite{McVittie, Klebanetal, Roshina, LakeAbdelqader, 
AndresRoshina}. Moreover, the old 
problem of the influence of the cosmic expansion on local dynamics (and {\em vice-versa}), 
which originally led to the study of such solutions  
\cite{McVittie}, is not completely 
solved \cite{CarreraGiulini}.

In this paper we analyze a Brans-Dicke solution found by Clifton, Mota, and Barrow 
\cite{CMB} and describing an inhomogeneity embedded in a FLRW universe. This solution is 
generated using a conformal transformation and the 
Husain-Martinez-Nu\~nez scalar field 
solution of general relativity \cite{HMN} 
as a seed. The conformal copy is not a perfect mirror image of 
the original solution, however, because it is found (in Sec.~\ref{section2}) 
that it describes a spacetime with 
properties quite different from the original one. This 
fact should not lead to superficial 
statements on the physical inequivalence between conformal frames because the usual conformal 
mapping between Brans-Dicke's and Einstein's theories (and their solutions) prescribes also 
a scaling of units in the Einstein frame  \cite{Dicke} 
which went lost in Ref.~\cite{CMB}, 
where the authors  
intended only to generate a new spherical and inhomogeneous solution of the Brans-Dicke 
field equations. Sec.~\ref{section3} contains a discussion on this subject.

\section{\label{section2}Understanding the Clifton-Mota-Barrow spacetime}

Clifton, Mota, and Barrow \cite{CMB} conformally mapped  the spherically symmetric and dynamical  
Husain-Martinez-Nu\~nez 
\cite{HMN} scalar field solution of general 
relativity to obtain the 2-parameter class of Brans-Dicke 
spacetimes 
\begin{widetext}
\be
ds^2 = 
-A^{\alpha \left( 1-\frac{1}{\sqrt{3}\, \beta}\right)} (r) 
\, dt^2
+A^{-\alpha \left( 1+\frac{1}{\sqrt{3}\, \beta}\right)} (r) 
\, t^{\frac{2 \left( \beta-\sqrt{3}\right)}{3\beta 
- \sqrt{3}} }  \left[ dr^2 
+ r^2 A(r)  d\Omega_{(2)}^2 \right] \,,\label{CMBmetric}
\ee 
\end{widetext}
\be
\phi( t,r) = A^{\frac{\pm 1}{2\beta}} (r)\, 
t^{\frac{2}{\sqrt{3}\,  \beta -1}} \,, 
\label{CMBscalarfield}
\ee
where $d\Omega_{(2)}^2=d\theta^2 +\sin^2 \theta \, d\varphi^2$ is the
 metric on the unit 2-sphere,
\begin{eqnarray}
A(r) & = & 1-\frac{2C}{r} \,,\\
&&\nonumber\\
\alpha &=& \pm \, \frac{\sqrt{3}}{2} \,,\\
&&\nonumber\\
\beta &=&\sqrt{2\omega+3} \,,
\end{eqnarray}
$C$ is a parameter related to the mass of the central inhomogeneity, and 
$\omega $ is the Brans-Dicke coupling parameter which is required to be 
larger than $-3/2$. We adopt the 
notations of Ref.~\cite{Wald}. There are spacetime singularities at $r=2C$ and 
at $t=0$, therefore, the relevant coordinate range is $2C<r<+\infty$ and 
$t>0$ \cite{CMB}. The scale factor of the spatially flat 
FLRW background universe is
\be \label{scalefactor}
a(t)= t^{\frac{\beta-\sqrt{3}}{3\beta-\sqrt{3}} }\equiv 
t^{\gamma} \,.
\ee
The line element~(\ref{CMBmetric}) can be rewritten as
\be
ds^2=-A^{\sigma}(r) \, dt^2 +A^{\Theta}(r) \, a^2(t)dr^2 +R^2(t,r)d\Omega_{(2)}^2 \,,
\label{lineelementq}
\ee
where
\begin{eqnarray}
\sigma &=& \alpha \left( 1-\frac{1}{\sqrt{3}\, \beta} \right) \,,\label{sigma}\\
&&\nonumber\\
\Theta &=& -\alpha \left( 1+\frac{1}{\sqrt{3}\, \beta} \right) \,,\label{Theta}
\end{eqnarray}
and 
\be\label{arealradius}
R(t,r)=A^{\frac{\Theta+1}{2}}(r) \, a(t)r
\ee
is the areal radius.

\begin{figure}[t]
\includegraphics[scale=0.35]{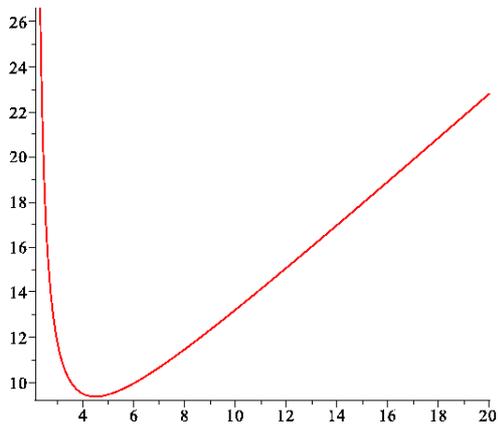}
\caption{\label{CMBfigure1} the areal radius $R$ (vertical 
axis) as a function of $r$ (horizontal axis) for 
the parameter values  $ \alpha=\sqrt{3}/2, \Theta=-3/2$, 
and $C=1$ (in units of $C$)  at a time $t_*$ at which 
$a(t_*)=1$.} 
\end{figure}

Let us examine the behaviour of the area $4\pi R^2$ of 
2-spheres of symmetry by studying how the areal radius 
behaves as a function of $r$.  We have
\begin{eqnarray}
\frac{\partial R}{\partial r} 
&=& a(t)A^{\frac{\Theta-1}{2}}(r)  
\left[  1-\frac{2C}{r} \, \frac{\left(1-\Theta \right)}{2}  
\right] \nonumber\\
&&\nonumber\\
&\,& \equiv  a(t)A^{\frac{\Theta-1}{2}}(r) 
\left( 1- \frac{r_0}{r} \right) \,,\label{partialR}
\end{eqnarray}
where 
\be
r_0=(1-\Theta)C
\ee
or, in terms of the proper (areal) radius,
\be \label{RAH}
R_0(t)=\left( \frac{\Theta+1}{\Theta-1} 
\right)^{\frac{\Theta+1}{2}}  (1-\Theta) a(t) \, C \,.
\ee
The critical value $r_0$  exists in the relevant spacetime 
region  $r_0>2C$ if $\Theta<-1$. 
In this case the areal radius can be written as 
\be
R(r)=\frac{ ra(t) }{ \left(1-\frac{2C}{r} 
\right)^{|\frac{\Theta+1}{2}|}}
\ee
and $R(r) \rightarrow +\infty$ as $r\rightarrow 2C^{+}$: 
the area of 2-spheres of symmetry 
diverges as $r\rightarrow 2C^{+}$. Due to 
eq.~(\ref{partialR}),  for $\Theta<-1$ we have 
$\frac{1-\Theta}{2}>1$ and $ 
\partial R/\partial r>0$ for $r$ larger than 
\be
r_0 = 2C \left( \frac{1-\Theta}{2} \right)> 2C \,,
\ee
$ \partial R/\partial r=0$ at $r=r_0$, and $\partial 
R/\partial r <0 $ for $r<r_0$. 
The function $R(r)$ has  a minimum at $r_0$  
(fig.~\ref{CMBfigure1}).      
The area of 2-spheres of symmetry decreases between $2C$ 
and $r_0$, where  it is minimum, then it increases again. 
There is a wormhole throat joining two 
spacetime regions (cf. Ref.~\cite{Hayward} for
a detailed wormhole theory). Note that, since 
\be 
\Theta= \mp \frac{\sqrt{3}}{2} \left( 1+ \frac{1}{\sqrt{3} \sqrt{2\omega+3} } \right) 
\ee 
for $\alpha=\pm \sqrt{3}/2$, the condition $\Theta<-1$ 
requires $\alpha =+\sqrt{3}/2$. 
This value of $\alpha$, however, is a necessary but not a 
sufficient condition for the apparent horizon to exist. The 
sufficient condition $\Theta<-1$ imposes the constraint on 
the Brans-Dicke parameter 
\be 
\omega < \frac{1}{2} \left[ \frac{1}{ \left( 2-\sqrt{3} 
\right)^2}-3 \right] \equiv \omega_0 \,, 
\ee 
which can be satisfied in the allowed 
parameter region $\omega>-3/2$. Therefore, for $-3/2< 
\omega <\omega_0 $ there is an apparent horizon at 
 $r_0>2C$ and the solution can be taken to represent a 
Brans-Dicke wormhole.  Here by ``wormhole'' we simply refer 
to a spacetime containing a smooth wormhole throat 
connecting  two spacetime regions. Other definitions 
of wormhole  (for example, a generalization to the 
dynamical case of  the definition of Ref.~\cite{Bronnikov}) 
are  more stringent and would not allow this spacetime to 
be called a wormhole.

The region $2C<r<r_0$ is not a FLRW 
region and the scalar field (\ref{CMBscalarfield}) is 
finite and 
non-zero at $r_0$: 
\be 
\phi\left(t, r_0 \right)= t^{ \frac{2}{\sqrt[]{3}\beta -1 } } \left( 
\frac{\Theta+1}{\Theta-1} \right)^{\frac{\pm 1}{2\beta} } 
\,.
\ee 
The proper radius  (\ref{RAH}) of the wormhole throat   is 
exactly comoving with the cosmic  substratum
 and disappears if the central 
inhomogeneity is removed, which is formally 
described by the limit $C\rightarrow 0$.

Let us now investigate the presence of apparent horizons in 
the 
metric~(\ref{CMBmetric}). Using eq.~(\ref{arealradius}) and 
substituting the relation between differentials
\be
dr=\frac{ dR-A^{\frac{\Theta+1}{2}}(r) \, \dot{a}(t)rdt}{
A^{\frac{\Theta-1}{2}}a(t) \, \frac{C( \Theta+1)}{r} 
+A^{\frac{\Theta+1}{2}}(r) \, a(t)}
\,,
\ee
into the line element (\ref{lineelementq}), one obtains
\begin{eqnarray}
ds^2 &=& -A^{\sigma}dt^2 +\left[ 
\frac{dR^2-2A^{\frac{\Theta+1}{2}} r\dot{a} \, dtdR +A^{\frac{\Theta+1}{2}} 
r^2 \dot{a}^2 dt^2}{ D_1(r)} \right] \nonumber\\
&\nonumber\\
&\, & +R^2 d\Omega_{(2)}^2 \,,
\end{eqnarray}
where
\be
D_1(r)=A(r) \left[ 1+\frac{C(\Theta+1)}{rA(r)} \right]^2 
\,.
\ee
Collecting similar terms yields
\begin{eqnarray}
ds^2 &=& - \frac{ \left( D_1A^{\sigma} -H^2R^2\right)}{D_1} 
\, dt^2 -\frac{2HR}{D_1}\, dtdR +\frac{dR^2}{D_1} 
\nonumber\\
&&\nonumber\\ 
&\, & + R^2  d\Omega_{(2)}^2 \,,
\label{eq:Delta}
\end{eqnarray}
where $H \equiv \dot{a}/a$ is the Hubble parameter of the 
background universe. The inverse of the metric $g_{\mu\nu}$ 
of eq.~(\ref{eq:Delta}) is
\be
\left( g^{\mu\nu} \right)=\left(
\begin{array}{cccc}
-\frac{1}{A^{\sigma}} & -\frac{HR}{A^{\sigma}} & 0 & 0 \\
&&&\\
-\frac{HR}{A^{\sigma}} & \frac{\left( D_1 
A^{\sigma}-H^2R^2\right)}{A^{\sigma}} & 0 & 0  \\
&&&\\
0 & 0 & \frac{1}{R^2} & 0 \\
&&&\\
0 & 0 & 0 &\frac{1}{R^2\sin^2\theta} 
\end{array} \right).
\ee
In the presence of spherical symmetry the apparent  
horizons are located by the roots of the equation  
$\nabla^cR\nabla_cR=0$ ({\em e.g.}, \cite{NielsenVisser}) 
or $g^{RR}=0$, which here yields
\be\label{locatingAH}
D_1 (r) A(r)= H^2(t) R^2(t,r) \,.
\ee
The left hand side of this equation depends only on $r$ 
while the right hand side depends on both $r$ and $t$. This 
equation can only be satisfied when the right hand 
side is 
time-independent  and the only possibility for this to 
occur is when $H=\gamma/t =0$, corresponding to $\gamma=0$, 
$\beta=\sqrt{3}$, and $\omega=0$. This value of the 
Brans-Dicke parameter   gives a static solution  describing 
a spherical inhomogeneity in a Minkowski background, which 
is discussed in the next section. With this exception, 
eq.~(\ref{locatingAH}) has no solutions and there are  no  
apparent horizons in the spacetime~(\ref{CMBmetric}). In 
particular, for  $\Theta<-1$ the wormhole throat is not an 
apparent horizon.

Let us discuss now the case $\omega \geq \omega_0$ and the 
case $\alpha=-\sqrt{3}/2$. In these  situations there is no 
wormhole throat and no apparent horizon in the $r>2C$ 
region and  the  Clifton-Mota-Barrow spacetime contains a 
naked singularity.

For $\alpha=-\sqrt{3}/2$ it is $ \Theta =\frac{\sqrt{3}}{2}  
\left( 1+\frac{1}{\sqrt{3}\, \beta} \right)>0$ and 
\be 
R(r)=\left( 1-\frac{2C}{r} \right)^{\frac{|\Theta+1|}{2}} a(t) r 
\ee 
goes to  zero as $r\rightarrow 2C^{+}$. Since $r_0<2C$, the 
areal radius $R(r)$ is always an increasing  function of 
$r$ in the relevant range $2C<r<+\infty$. This spacetime 
contains a naked  singularity at $R=0$.

\section{\label{}The special case $\omega=0$}

The value $\omega=0$ of the Brans-Dicke coupling, 
corresponding to $\beta=\sqrt{3}$ and $\gamma=0$, produces 
the static metric
\be\label{StaticMetric}
ds^2=-A^{\frac{2\alpha}{3}}(r) dt^2 
+\frac{dr^2}{A^{\frac{4\alpha}{3}}(r)} 
+\frac{r^2}{A^{\frac{4\alpha}{3}-1}(r)}\, d\Omega_{(2)}^2 
\ee
and the scalar field  
\be\label{StaticScalar}
\phi(t,r)=A^{ \frac{\pm 1}{2\sqrt{3}}}(r) t \,,
\ee
which is time-dependent even though the metric is 
static.\footnote{This is not the only occurrence of this 
circumstance: a similar situation is known  for 
the static limit of another separable solution of the 
Brans-Dicke field equations found by Clifton, Mota, and 
Barrow \cite{FVSL}.} 

The metric~(\ref{StaticMetric}) is easily identified as a 
member of the Campanelli-Lousto class 
\cite{CampanelliLousto}. The general Campanelli-Lousto 
solution has the form
\begin{eqnarray}
ds^2 &=& -A^{b+1}(r)dt^2+\frac{dr^2}{A^{a+1}(r)}+ \frac{r^2 
d\Omega_{(2)}^2}{A^a(r)} \,, \label{CLmetric}\\
&&\nonumber\\
\phi(r) & = & \phi_0 A^{\frac{a-b}{2}}(r) \,, 
\label{CLscalar}
\end{eqnarray}
and $\phi_0, a$, and $b$ are constants with $\phi_0>0$. The 
Brans-Dicke parameter is given by \cite{CampanelliLousto}
\be
\omega(a,b)= -2\, \frac{\left( 
a^2+b^2-ab+a+b\right)}{\left( a-b \right)^2} 
\,.\label{omegaCL}
\ee 
In the case of the metric~(\ref{StaticMetric}) setting
\be
\left( a, b \right)= \left(  \frac{4\alpha}{3} -1, 
\frac{2\alpha}{3} -1 \right) 
\ee
reproduces the Campanelli-Lousto metric~(\ref{CLmetric}).  
Then, the expression~(\ref{omegaCL})
 gives $\omega \left( \frac{4\alpha}{3}-1,  
\frac{2\alpha}{3}-1 \right)=0$ for $\alpha =\pm 
\sqrt{3}/2$. However, the scalar field~(\ref{StaticScalar}) 
differs from the Campanelli-Lousto scalar~(\ref{CLscalar}) 
by the linear dependence on the time $t$. Thus, the static 
limit of the Clifton-Mota-Barrow solution provides a 
(rather trivial) generalization of a Campanelli-Lousto 
solution.

The nature of the Campanelli-Lousto spacetime depends 
on the sign of the parameter $a$ \cite{VZF} which, in our 
case, corresponds to the choice $\alpha=\sqrt{3}/2$ or 
$-\sqrt{3}/2$.  For $a \geq 0$ (corresponding to 
$\alpha=+\sqrt{3}/2,  a \simeq 0.1547$, and $\Theta 
=-\frac{4\alpha}{3} \simeq -1.1547 <-1$) 
the 
Campanelli-Lousto spacetime contains a wormhole  
throat coinciding with an apparent horizon and located at 
$r_0=2C \left( \frac{1-\Theta}{2}\right)>2C$ \cite{VZF}. 
This is consistent with eq.~(\ref{locatingAH}) with $H=0$ 
since, in this case, the equation $g^{RR}=0$ locating the 
apparent horizons reduces to $D_1(r)=0$ which yields again 
the root $r_0=C(1-\Theta)$ lying in the physical 
region $r>2C$.\footnote{The discussion of 
Ref.~\cite{VZF}, however, does not depend on setting 
our $D_1=0$.}  This is the only case in which the  
Clifton-Mota-Barrow solution under study contains an 
apparent horizon.

For $a<0$ (which is reproduced by the choice 
$\alpha=-\sqrt{3}/2$ and gives $a\simeq -2.1547$ and 
$\Theta \simeq 1.1547>0$) there are no apparent horizons 
and the spacetime 
contains a naked singularity \cite{VZF}. This is 
consistent with the fact that eq.~(\ref{locatingAH}) 
with $H=0$ can only be satisfied if $D_1(r)=0$ and in this 
case there are no acceptable solutions because $r_0<2C$.

\section{\label{section3}Discussion and conclusions}

According to the parameter values, the Clifton-Mota-Barrow 
spacetime (\ref{CMBmetric}) contains a wormhole or a naked 
singularity (black holes, wormholes, and naked  
singularities could in principle be distinguished 
observationally through gravitational 
lensing \cite{lensing}). 
 In the last situation, this solution  of the Brans-Dicke 
field equations cannot be obtained as the development of 
regular Cauchy data.

One question which arises is the following: the Husain-Martinez-Nu\~nez 
and the Clifton-Mota-Barrow spacetimes are conformally related. As explained long ago 
by Dicke \cite{Dicke}, the Jordan and 
the Einstein conformal frames should be different representations of the same physics (provided 
that the conformal transformation does not break down)---this issue has been the subject 
of a lively debate but has been shown to be largely a pseudo-problem (see \cite{Flanagan,
FaraoniNadeau, DeruelleSasaki} and the references therein). Then, why does the same solution look 
so different in the two different conformal frames for the parameter values for which a 
Jordan frame wormhole or naked singularity (\ref{CMBmetric}) 
corresponds to the Einstein frame black hole of 
\cite{HMN}? The answer is that, by following the more ordinary route and 
conformally transforming the Clifton-Mota-Barrow metric and scalar field (\ref{CMBmetric}) 
to the Einstein frame would produce the Husain-Martinez-Nu\~nez 
metric {\em with scaling units} of length, time and mass. What is physically relevant is the 
ratio of a physical quantity to its unit, and the units change with the spacetime position.  
Specifically, the units of length and time scale as the conformal factor $\Omega$, while the 
unit of mass scales as $\Omega^{-1}$ and derived units scale 
accordingly \cite{Dicke}.  In the Einstein frame, matter is 
coupled non-minimally to the metric while in the Jordan 
frame matter is minimally coupled.  The scaling of units 
in the Einstein frame goes hand  in hand with the 
non-minimal coupling of matter to the metric. {\em In 
vacuo} (which is the situation contemplated here), the 
non-minimal coupling of matter is forgotten, but the 
scaling of units should be remembered.

Another issue is that, contrary to event horizons (which are null surfaces and are 
conformally invariant), apparent horizons (which can be spacelike or even timelike) are not 
conformally invariant and change location under a conformal transformation \cite{AlexValerio}. 
In order to characterize the properties of a dynamical black hole when conformal transformations
are involved, one should not consider the apparent horizons of a metric but a new surface 
characterized by an entropy 2-form, as explained in detail 
in  \cite{AlexValerio} and 
\cite{AlexFirouzjaee}. The new prescription of  
\cite{AlexValerio} takes into account 
the scaling of units in the Einstein frame. 

Therefore, a metric obtained from the conformal transformation to the Einstein frame of a 
seed Jordan frame metric with the extra information that units are scaling is quite 
different from the same formal metric with fixed units, which explains why conformally 
related spacetimes can look very different. Clifton, Mota, and Barrow took the 
Husain-Martinez-Nu\~nez solution of general relativity {\em with fixed units} and used it 
as a seed to generate a new class of solutions of Brans-Dicke gravity---they did not 
worry about generating a {\em physically equivalent} solution, which would have required 
to take into account scaling units. This procedure is 
certainly legitimate and achieves 
the goal, but it generates physically inequivalent 
spacetimes when the requirement of scaling units is 
dropped. Indeed, there are comments in the literature about 
the fact  that conformally related solutions of Brans-Dicke 
theory and of the Einstein equations  do not share 
the same properties \cite{coldBHs}. A similar situation 
occurs with the Campanelli-Lousto  solutions of Brans-Dicke 
theory \cite{CampanelliLousto}, which relate to 
Fisher-Janis-Newman-Winicour solutions of the Einstein 
equations in the Einstein frame  \cite{VZF}, and with the 
veiled black holes of \cite{DeruelleSasaki, AlexValerio, 
AlexFirouzjaee} (see \cite{VZF} for a detailed discussion). 
To conclude, the physical  nature of the 
Clifton-Mota-Barrow class of solutions is now clear  
and they  do not  cause problems for the interpretation of 
conformal frames.

\begin{acknowledgments}
We thank Prof. Wolfgang Graf for pointing out a mistake and  
typographical errors in a previous version of this 
manuscript.  This work is supported by  Bishop's University 
and by the Natural Sciences and 
Engineering  Research Council of Canada. 
\end{acknowledgments}




\end{document}